\documentclass{jetpl}
\twocolumn
\usepackage{color}
\usepackage{mathrsfs}
\usepackage{graphicx}
\usepackage{amsmath}
\usepackage{ulem}

\lat

\title{ Reply to comment on "Noise in the helical  edge channel  anisotropically coupled to a local spin"}
\rtitle{}
\sodtitle{}
\author{K.\,E. Nagaev$^{+}$\thanks{e-mail: nag@cplire.ru}, S.\,V. Remizov$^{+*}$,\,and\,D.\,S. Shapiro$^{+*}$}
\rauthor{}
\sodauthor{}
\address{+ Kotelnikov Institute of Radioengineering and Electronics, Mokhovaya 11-7, Moscow 125009, Russia}
\address{* Dukhov Research Institute of Automatics (VNIIA), Moscow 127055, Russia}

\dates{\today}{*}

\begin{document}

\maketitle

\newcommand{\bn}{{\bf n}}
\newcommand{\bp}{{\bf p}}   
\newcommand{\br}{{\bf r}}
\newcommand{\bk}{{\bf k}}
\newcommand{\bv}{{\bf v}}
\newcommand{\brho}{{\bm{\rho}}}
\newcommand{\bj}{{\bf j}}
\newcommand{\wk}{\omega_{\bf k}}
\newcommand{\nk}{n_{\bf k}}
\newcommand{\eps}{\varepsilon}
\newcommand{\la}{\langle}
\newcommand{\ra}{\rangle}
\newcommand{\be}{\begin{eqnarray}}
\newcommand{\ee}{\end{eqnarray}}
\newcommand{\intl}{\int\limits_{-\infty}^{\infty}}
\newcommand{\dE}{\delta{\cal E}^{ext}}
\newcommand{\SE}{S_{\cal E}^{ext}}
\newcommand{\dsp}{\displaystyle}
\newcommand{\phit}{\varphi_{\tau}}
\newcommand{\p}{\varphi}
\newcommand{\dphi}{\delta\varphi}
\newcommand{\dbj}{\delta{\bf j}}
\newcommand{\dI}{\delta I}
\newcommand{\dph}{\delta\varphi}
\newcommand{\ua}{\uparrow}
\newcommand{\da}{\downarrow}
\newcommand{\sv}{\sigma_{\alpha}v_F}
\newcommand{\cotV}{\coth\!\left(\frac{e V}{2 T}\right)}

The authors of comment \cite{comment19} claim  that our recent results \cite{Nagaev18} on the noise in 
the helical edge channel of a 2D topological insulator coupled to a spin-1/2 impurity are incorrect. Their argument is that the
expression for the average backscattering current that follows from our Eq. (7) differs from Eq. (22) of their own  
paper \cite{Kurilovich17}. They state that it is illegal to assume that the density matrix of 
the impurity spin is diagonal in the basis of $S_z$, which is the cornerstone of our calculations and the calculations of a 
previous paper \cite{Kimme16}. 

The authors of the comment reason that the dephasing of the impurity spin arises not only from 
the term $J_z S_z s_z$ in the Hamiltonian, but also  from the term $2J_a S_x s_z$. However this depends on the relative  
magnitude of the parameters $J_z$ and $J_a$. In our paper we clearly state that the dephasing of the impurity spin is due to 
the term $J_z S_z s_z$, and this implies that $J_z$ is large. This does not mean that the exchange matrix is diagonal 
{as stated in \cite{comment19},} but only 
means {that $J_{33}$ in Eq. (1) of the comment is much larger than all the other elements of the matrix}. 
Note that this parameter does not enter into any of the 
transition rates $\Gamma_0^{\pm}$, $\Gamma_a$, $\Gamma_1$, or $\Gamma_2$ and its large value does not impose any restrictions 
on the relations between these quantities. 

The authors of the comment admit that in this approximation, their Eq. (5) crosses 
over to Eq. (7) of our 
paper Ref. \cite{Nagaev18} written for $J_2=J_a=0$. In addition, it is clearly seen that the voltage-proportional current in the limit of $J_2 = J_1 =0$ and $eV \gg T$, Eq.~(3) of the comment, vanishes in this case. Hence there is no contradiction between papers \cite{Nagaev18} and \cite{Kurilovich17}.

To summarize, our results are correct within the limits of applicability of our model, and their critique by the authors of the comment is irrelevant.

\end{document}